\documentclass[%
prresearch,
twocolumn,
superscriptaddress,
 amsmath,amssymb,
 aps,
]{revtex4-2}

\usepackage{graphicx}% Include figure files
\usepackage{dcolumn}% Align table columns on decimal point
\usepackage{bm}% bold math
\usepackage{amsmath}
\usepackage{xcolor}
\graphicspath{ {./figs/} }
\usepackage{placeins}
\usepackage{algpseudocode}
\usepackage[]{changes}
\definechangesauthor[name={Indrajit's remarks}, color=blue]{IT}
\definechangesauthor[name={Sadjad}, color=blue]{SA}

\makeatletter
\newcommand*{\rom}[1]{\expandafter\@slowromancap\romannumeral #1@}
\makeatother

\begin{document}

% \preprint{APS/123-QED}
\title{Rigidity of Epithelial Tissues as a Double Optimization Problem}

\author{Sadjad Arzash}
 \affiliation{Department of Physics, Syracuse University, Syracuse, NY 13244, USA}
 \affiliation{Department of Physics and Astronomy, University of Pennsylvania, Philadelphia, PA 19104, USA}
\author{Indrajit Tah}
\affiliation{Speciality Glass Division, CSIR-Central Glass and Ceramic Research Institute, Kolkata 700032, India}
\affiliation{Academy of Scientific and Innovative Research (AcSIR), Ghaziabad 201 002, India}
\author{Andrea J. Liu}
\affiliation{Department of Physics and Astronomy, University of Pennsylvania, Philadelphia, PA 19104, USA}
\author{M. Lisa Manning}
\affiliation{Department of Physics, Syracuse University, Syracuse, NY 13244, USA}

\begin{abstract}
How do cells tune emergent properties at the scale of tissues? One class of such emergent behaviors are rigidity transitions, in which a tissue changes from a solid-like to a fluid-like state or vice versa. Here, we introduce a new way for a tissue described by a vertex model to tune its rigidity, by using ``tunable degrees of freedom." We use the vertex model elastic energy as a cost function and the cell stiffnesses, target shapes, and target areas as different sets of degrees of freedom describing cell-cell interactions that can be tuned to minimize the cost function. We show that the rigidity transition is unaffected when cell stiffnesses are treated as tunable degrees of freedom. When preferred shapes or areas are treated as tunable degrees of freedom, however, induced spatial correlations in target cell shapes or areas shift the rigidity transition. These observations suggest that tissues can coordinate changes in cell-scale properties, treated here as tunable degrees of freedom, to achieve desired tissue-scale behaviors.
\end{abstract}

\maketitle

\section{Introduction}

The molecular processes that govern the formation of biological tissues operate at the cellular level but give rise to collective behavior at the multicellular scale. Similarly, in systems such as mechanical, flow or electrical networks instructions encoded in the microscopic structure control collective properties. In materials design, the process of achieving a specific functionality typically involves a series of iterative steps in which the system is continually tested for desired functionality, adjusted based on feedback, and tested again to refine its performance. An effective strategy for solving this inverse problem of material design in these systems is gradient descent on a \emph{cost function} that embodies the desired collective property by tuning microscopic \emph{tunable degrees of freedom} (DOFs) characterizing interactions, such as the presence or absence of a bond~\cite{goodrich_principle_2015,rocks_designing_2017,hexner_role_2018, hexner_linking_2018}, bond stiffnesses~\cite{rocks_limits_2019,pashine_directed_2019} or rest lengths in elastic networks, or conductances~\cite{rocks_limits_2019} in flow or electrical networks. Physics dictates that each system must also satisfy physical constraints during this process, imposed by minimizing the energy in elastic networks or dissipated power in flow or electrical networks, with respect to \emph{physical degrees of freedom} (node positions in elastic networks, node pressures/voltages in flow/electrical networks).

Simultaneous minimization of the cost function and energy/power with respect to tunable and physical degrees of freedom (\emph{double optimization}) can be used to generate an auxetic~\cite{goodrich_principle_2015,hexner_role_2018, hexner_linking_2018} or allosteric response~\cite{rocks_designing_2017, rocks_limits_2019}. Alternatively, minimization of the energy/power while varying tunable degrees of freedom according to local rules~\cite{stern_learning_2023} can also be effective. Such local update rules include those that naturally occur in real materials, like directed aging~\cite{pashine_directed_2019,hexner_effect_2020,hexner_periodic_2020}, as well as rules that approximate gradient descent, as in Equilibrium Propagation~\cite{scellier_equilibrium_2017} or Coupled Learning~\cite{stern_supervised_2021}. These ideas have led to successful learning of desired properties in the lab~\cite{rocks_designing_2017,pashine_directed_2019,hexner_effect_2020,dillavou_demonstration_2022,wycoff_desynchronous_2022, dillavou_machine_2023}. 

Here we show that biological tissues can potentially tune cell-scale properties, viewed as tunable DOFs, to drive robust macroscopic, collective behaviors necessary for development and evolution. Our work focuses on rigidity transitions, which are a specific example of macroscopic collective behavior. Rigidity transitions occur when the tissue collectively switches back and forth from fluid-like behavior, where cells are able to rearrange neighbors and the tissue can accommodate significant strain, to a solid-like behavior, where cells do not change neighbors and straining the tissue costs energy. Recent experiments demonstrate that tissues shift from a solid to a fluid~\cite{mongera_fluid--solid_2018,wang_anisotropy_2020} or near-fluid state~\cite{claussen_geometric_2023} as a function of space~\cite{mongera_fluid--solid_2018} and time~\cite{wang_anisotropy_2020}, to facilitate flows necessary for body axis elongation~\cite{mongera_fluid--solid_2018, wang_anisotropy_2020,claussen_geometric_2023} and organ formation~\cite{erdemci-tandogan_tissue_2018, sanematsu_3d_2021}. A well-vetted class of simple biophysical models (vertex~\cite{hufnagel_mechanism_2007, farhadifar_influence_2007, park_unjamming_2015,bi_density-independent_2015}, Voronoi~\cite{bi_motility-driven_2016}, and cellular Potts~\cite{chiang_glass_2016} models) have successfully made quantitative predictions -- with no fit parameters -- for rigidity transitions in confluent epithelial tissues~\cite{wang_anisotropy_2020, park_unjamming_2015, devany_cell_2021}. A key feature of vertex models, validated in experiments, is that the rigidity transition is controlled by a geometric cell shape factor. This shape factor serves as a coarse-grained parameter that encapsulates the effects of molecular-scale processes, such as contractility driven by myosin and adhesion regulated by E-cadherin \cite{sahu_small-scale_2020, wang_e-cadherin_2024}.

In this paper, we explore the idea that developmental processes can usefully be regarded as double optimization processes, in which cell-scale tunable DOF, such as cell shape, are adjusted to optimize a tissue-scale cost function that is minimized when the tissue achieves a desired macroscopic final state, while simultaneously staying in mechanical equilibrium.
%generate observed tissue-scale behaviors.
This viewpoint is bolstered by the recent finding that the \textit{Drosophila} amnioserosa appears to shift its rigidity transition to remain rigid throughout the developmental process of dorsal closure by tuning preferred cell shapes continuously throughout the process~\cite{tah_minimal_2025}. By framing a developmental process as a double optimization problem, we can unambiguously identify which cell-scale parameters within a vertex model are important for controlling a given macroscopic property. We argue that double optimization represents a theoretical framework for identifying cell-scale and molecular mechanisms that control larger-scale behavior, which is a major open problem in cell and developmental biology. This framework allows us to study an \emph{ensemble} of tissue states that all minimize the same cost function. If we can identify common features in this ensemble that emerge from the double optimization process, we can then search for such features in biological experiments.

%answering the referee comment on "why this problem is interesting from a physics perspective?" 
This problem is also interesting from a physics perspective. Previous work has focused on over-constrained networks or jammed packings, in which the parameter that controls rigidity is the coordination number that describes the number of constraints per particle or node. This is because such systems become rigid when the number of degrees of freedom equals the number of constraints. Work by Hagh, et al.~\cite{hagh_transient_2022} introduced tunable DOFs in the form of particle radii, and showed that these can be used to control rigidity over a wide range by tuning the coordination number, enabling the design of highly stable jammed states~\cite{hagh_transient_2022}.

In contrast, vertex models are highly under-constrained, i.e., the number of physical degrees of freedom (vertex positions) is much larger than the number of constraints.  Vertex models become rigid through geometric incompatibility, where cells are unable to achieve their target perimeters and areas. The system is stabilized due to energetic costs that occur only at second order in perturbations to the constraints~\cite{damavandi_energetic_2022}, the same mechanism that drives strain-induced rigidity in sub-isostatic fiber networks \cite{sharma_strain-controlled_2016, shivers_scaling_2019, merkel_minimal-length_2019, arzash_finite_2020}. This raises the question of whether rigidity can be controlled in vertex models using tunable DOFs.

As a first step towards addressing these open physics and biology questions, we investigate tunable DOFs in 2D vertex models. We study how different sets of allowed tunable DOFs -- specifically, cell stiffnesses, preferred areas, or preferred perimeters --affect our ability to minimize a cost function.  Here, as proof of principle, we make the simplest possible choice, analogous to Ref.~\cite{hagh_transient_2022} for the cost function jammed packings: the total mechanical energy of the system. In other words, we explore the ability of different sets of tunable DOFs to drive the system towards zero-energy floppy/fluidized states. 

To characterize the sensitivity of vertex models to tunable DOFs, we must also account for an important consideration. Such models can be driven towards a fluid-like state by simply altering the mean~\cite{bi_density-independent_2015} or the width~\cite{li_mechanical_2019} of the distribution of cell shapes. A similar result was discovered in over-constrained jammed packings, where rigidity was found to be trivially dependent on the first and second moments of the radii distribution~\cite{hagh_transient_2022}. As in that previous work~\cite{hagh_transient_2022}, we avoid these trivial dependencies by fixing the distribution (or a set of its moments) and asking whether double optimization is able to introduce \emph{spatial correlations} in the tunable DOFs that are sufficient to shift the rigidity transition. If the system is able to learn, our next goal is to identify which tunable DOFs are able to control the rigidity transition, and identify observable features that distinguish states that have learned from those that have not.

\begin{figure}
	\includegraphics[width=8cm,height=8cm,keepaspectratio]{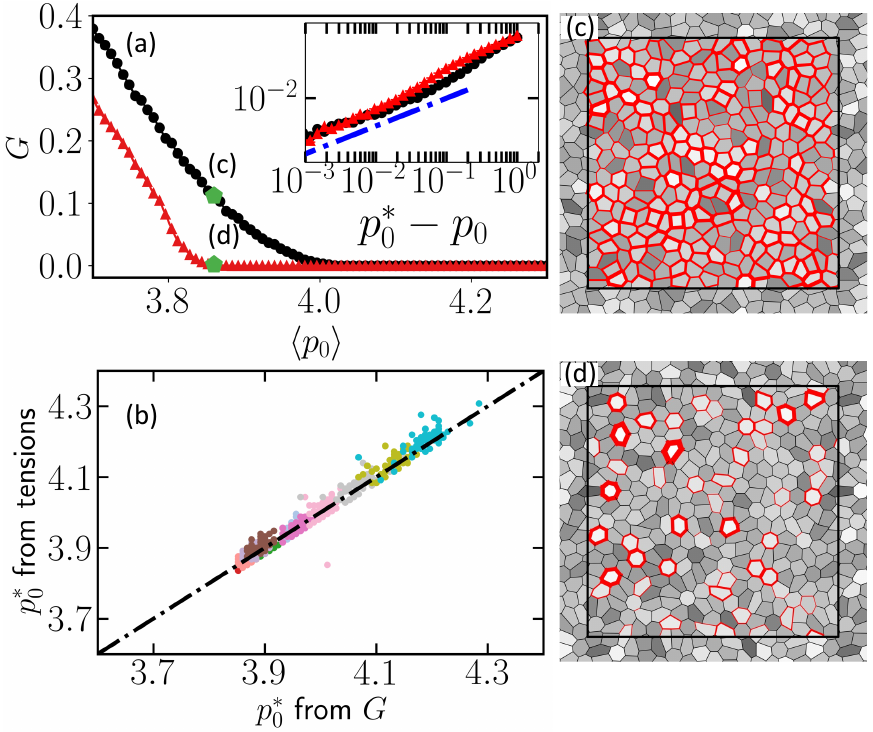}
	\caption{\label{fig:model_images} (a) Shear modulus $G$ versus average target shape $\langle p_0 \rangle$ in vertex models with polydisperse $p_{0i}$. Black curve (circles) shows minimization based solely on physical degrees of freedom, while red curve (triangles) includes both physical and $\{p_{0i}\}$ degrees of freedom. Inset illustrates shear modulus scaling; dashed blue line indicates a slope of $1.0$. (b) Rigidity transition point $p_0^*$ from edge tension percolation versus shear modulus $G$ with different $\sigma$ values for $\{p_{0,i}\}$ as degrees of freedom (DOF). Black dashed line represents $y = x$. (c, d) Tissue structures for highlighted points in (a). Cells are colored based on their $p_{0,i}$ values (higher $p_{0,i}$ is darker). Edge tensions are shown in red, with thickness proportional to tension. Both snapshots have the same distribution of target shape factors $\{p_{0,i}\}$.}
	
\end{figure}

\section{Model}

We study a 2D vertex model~\cite{honda_description_1978, farhadifar_influence_2007}, which describes a tissue monolayer as a network of polygonal cells. The physical DOF are the polygon vertices. Cellular properties and interactions are encoded in an energy function \(E = \sum_{i}^{N} \left[ K_{A,i}(A_i - A_{0,i})^2 + K_{P,i} (P_i - P_{0,i})^2 \right]\), where $A_i$ and $A_{0,i}$ are the actual and preferred areas, $P_i$ and $P_{0,i}$ are the actual and preferred perimeters, $K_{A,i}$ and $K_{P,i}$ are the area and perimeter moduli of cell $i$. It is helpful to make the above equation dimensionless using $\langle K_{A,i} \rangle \langle A_{0,i} \rangle^2$ as the units of energy and $\sqrt{\langle A_{0,i} \rangle }$ as the units of length. We then have
\begin{equation}
	\label{eq:energy_dimless}
	e = \sum_{i}^{N} \left[ k_{a,i}(a_i - a_{0,i})^2 + k_{p,i} (p_i - p_{0,i})^2 \right],   
\end{equation}
where $\langle k_{a,i} \rangle =1$, $\langle a_{0,i} \rangle = 1$, and $p_i$, $p_{0,i}$ are the dimensionless actual and preferred shape indices.
Eq.\ 1 has been well studied for the case where $k_{a,i}$, $a_{0,i}$, $k_{p,i}$ have delta-function distributions, and $p_{0,i}$ has a distribution of zero~\cite{farhadifar_influence_2007,bi_density-independent_2015, bi_motility-driven_2016} or nonzero width~\cite{li_mechanical_2019}. Here, we study Eq.~\ref{eq:energy_dimless} using the open-source CellGPU code \cite{sussman_cellgpu_2017}, promoting $k_{a,i}$, $a_{0,i}$, $k_{p,i}$ and $p_{0,i}$ to tunable DOF. Initially, $N$ cell centers are set by random sequential addition in a square box with length $L=\sqrt{N}$; vertices and edges are defined from a Voronoi tessellation of these points. The energy in Eq.\ 1 is minimized using the FIRE algorithm~\cite{bitzek_structural_2006}.

We investigate the impact of various sets of tunable DOF separately; e.g., when $p_{0,i}$ are tunable DOF, we initialize $p_{0i}$ values from a Gaussian distribution with mean $\langle p_{0i} \rangle$ and standard deviation $\sigma$, and set $k_{a,i} = k_{p,i} = a_{0,i} = 1$ for all cells. As in Hagh, et al.~\cite{hagh_transient_2022}, we focus on the case where the  cost function is simply the energy, or the physical cost function. Hagh et al. have shown that in sphere packings, minimizing the energy with respect to both physical DOF (particle positions) and tunable DOF (particle radii) allows the system to find very rare low-energy states~\cite{hagh_transient_2022}, shifting the jamming transition. We minimize the energy (Eq.\ 1) with respect to both physical DOF (vertex positions) and tunable DOF to study the influence of tunable DOF on the rigidity transition. We keep the tunable DOF distributions approximately fixed by imposing constraints on sets of moments of the distribution, such as the $m = \{-1, -2, -3, 1, 2, 3\}$ moments (see the SI). This constrained minimization method ensures the distribution of tunable DOF stays fixed during our minimization dynamics. We also perform zero-temperature swap minimization to fix the distribution exactly. In this method, each of the $N$ cells maintains its preferred property (introducing $N$ constraints), cells are swapped in a trial move, and moves that lower the energy are accepted (see SI).

%\added[id=IT] {please define the value of G e.g. $G=10^{-X}$ which you called 0} 
We evaluate rigidity based on the shear modulus $G$~\cite{merkel_geometrically_2018}: $G>0$ in the rigid phase, and $G=0$ in the fluid phase. To compute $G$, we freeze all tunable DOF (see SI).  Unless otherwise stated, error bars show the standard deviation over 50 samples.

\section{Results}

\subsection{Introducing target shapes and areas as tunable degrees of freedom can fluidize tissues}

As the preferred shape index $p_0$ increases, confluent tissues with only physical degrees of freedom experience a solid-fluid phase transition at a critical value $p_0^*$ \cite{bi_density-independent_2015,merkel_minimal-length_2019}. For systems with polydisperse $p_{0,i}$, the critical point $p_0^*$ shifts towards larger average preferred shape factors with increasing standard deviation $\sigma$ of the $p_0$ distribution~\cite{li_mechanical_2019} (black data in Fig. \ref{fig:varying_sigma}a). For a system with $p_{0,i}$ drawn from a Gaussian distribution with $\sigma=0.2$~\cite{wang_anisotropy_2020}, we find $p_0^* = 4.05 \pm 0.02$ (curve with black circles in Fig.\ \ref{fig:model_images}a). As $p_0 \rightarrow p_0^{*-}$, approaching the transition from the rigid side, the shear modulus vanishes as a power law: $G \approx a (p_0^* - p_0)^b$ with $b=1.0$ \cite{bi_density-independent_2015,merkel_minimal-length_2019}. We subtract a finite-size-effect offset (see  SI) and fit to this form to see how the scaling exponent $b$ and the position of the rigidity transition, $p_0^*$, change as we introduce different sets of tunable DOF.

Rigidity is associated with percolation of edges (cell-cell junctions) with nonzero tensions~\cite{li_mechanical_2019}. The tension of edge $ij$ separating cells $i$ and $j$ is \(T_{ij} = 2 K_{P,i} (P_i - P_{0,i}) + 2 K_{P,j} (P_j - P_{0,j})\), which when nondimensionalized becomes:
\begin{equation}
	t_{ij} = 2 k_{p,i} \sqrt{a_{0,i}} \tau^p_i + 2 k_{p,j} \sqrt{a_{0,j}} \tau^p_{j},  
	\label{eq:tension}
\end{equation}
where $\tau^p_i = p_i - p_{0,i}$ is the tension of cell $i$ in units of $\langle K_{A,i} \rangle \langle A_{0,i} \rangle^{3/2}$, i.e., energy/length. For
$p_0 < p_0^*$, a percolating cluster of nonzero edge tensions (Fig.\ \ref{fig:model_images}) maintains mechanical rigidity of tissue \cite{li_mechanical_2019}. For $p_0 > p_0^*$, nonzero edge tensions fail to percolate and the tissue is fluid -- it cannot resist shear deformation. 

\begin{figure}
	\includegraphics[width=8cm,height=8cm,keepaspectratio]{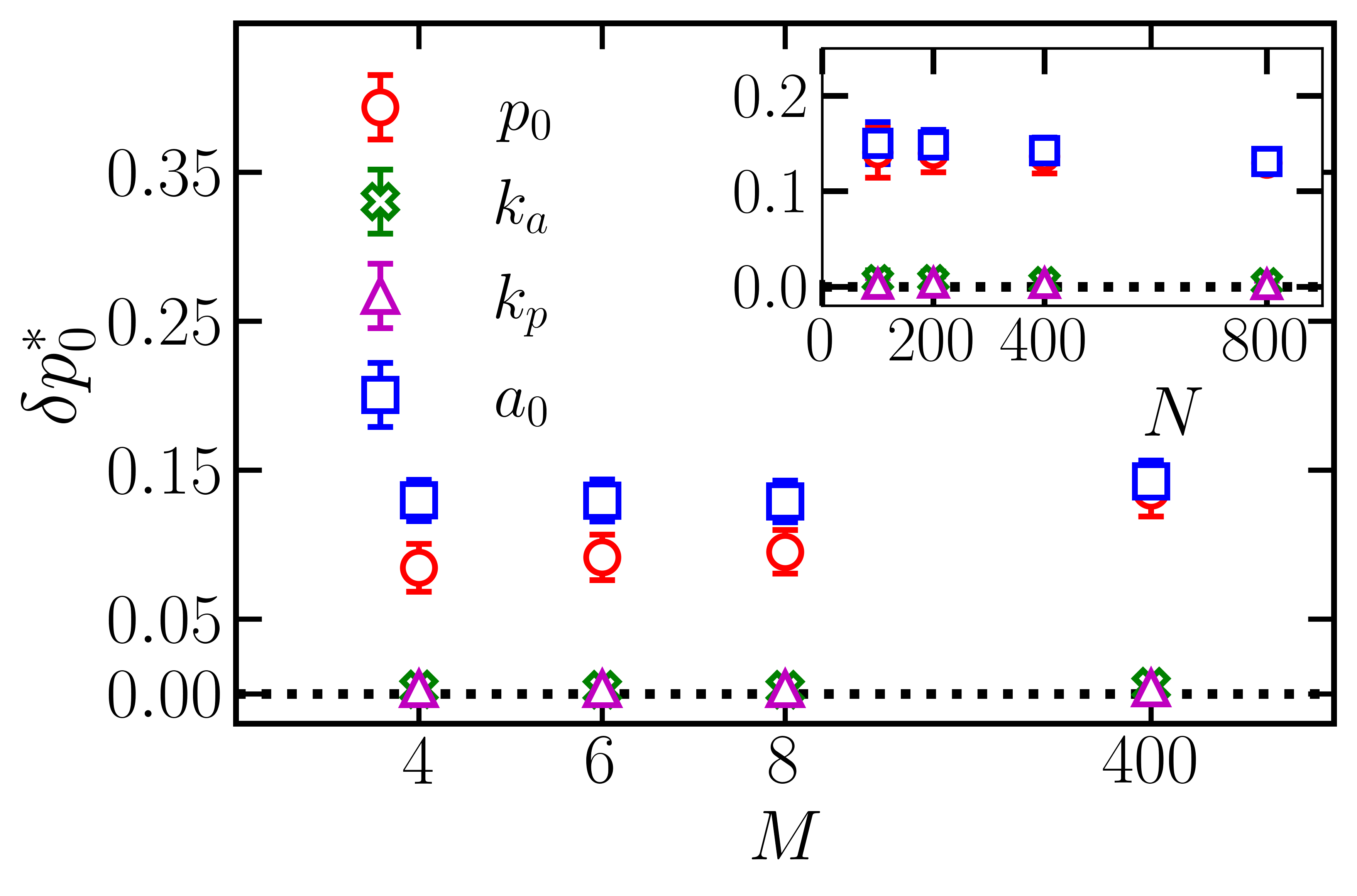}
	\caption{\label{fig:exponents} Change in the rigidity transition point $\delta p^*_0$ after introducing different transient degrees of freedom (different symbols), as a function of the number of moment constraints $M$ on the distribution. Specifically, the exact moments for \( M \) constraints are \( \{-M/2, ..., M/2\} \), excluding zero. Inset shows how $\delta p^*_0$  varies with $N$ for $M=400$. The zero-temperature swap system is indicated by 400 constraints, the number of cells in the tissue. These results correspond to a standard deviation of $\sigma=0.2$ of transient DOF.
	}
\end{figure}

We first note that $p_0^*$ is unaffected when the cell perimeter stiffnesses $\{k_{p,i}\}$ in Eq.~\ref{eq:energy_dimless} are allowed as tunable DOF (Fig. \ref{fig:exponents}). This observation is consistent with the fact that, in the case of uniform $\{p_{0,i}\}$, the deviations in perimeter $\tau_i^p = p_i - p_{0,i}$ are geometrically constrained to be non-negative in the solid phase, which prevents the stiffness degrees of freedom $\{k_{p,i}\}$ from altering the percolation of edge tensions. This result aligns with prior studies of jamming in sphere packings \cite{hagh_transient_2022}, where stiffness degrees of freedom are similarly irrelevant in shifting the transition point. While this observation is supported by our numerical results, a rigorous mathematical proof of this effect is nontrivial and is reserved for future investigation. The scaling exponent $b$ also remains unchanged but the tissue softens (see SI).

We next consider variations in cell area stiffnesses $k_{a,i}$ as tunable DOF with $a_{0,i}=1$, $k_{p,i}=1$ and $p_{0,i}=p_0$ for every cell $i$. One might expect the system to distribute its cell areas $a_i$ to be closer to $a_{0,i}=1$ for cells with larger values of $k_{a,i}$, leading to correlations that shift the transition. However, vertex models are unstressed at the rigidity transition \cite{damavandi_energetic_2022}, so their properties there cannot depend on $k_{a,i}$. As a result, the rigidity transition is unaffected by introducing $k_{a,i}$ as tunable DOF.

We now consider preferred shape indices $\{p_{0,i}\}$ as tunable DOF. Upon minimization, tissues adjust the values of some individual preferred shape indices $p_{0,i}$ to lower the energy by eliminating $p_i-p_{0,i}$. This leads to a lower fraction of nonzero tension edges, shifting the rigidity transition $p^*_{0}$ to lower values (red triangles in Fig. \ref{fig:model_images}a). Since typical shape indices observed in experiments range from about $3.8$ to $4.3$~\cite{park_unjamming_2015, wang_anisotropy_2020}, the shift in the transition point from about $3.85$ to about $4$ is quite significant. Thus, minimizing $E$ with respect to $p_{0,i}$ as well as the vertex positions introduces spatial correlations in $p_{0,i}$ that fluidize a tissue that would otherwise be solid. This shift persists whether we constrain certain moments of the $p_0$-distribution or preserve the distribution exactly (see Fig. \ref{fig:exponents}). The scaling exponent $b$ for the shear modulus remains unaffected within our error bars (see SI). Moreover, the amplitude $a$ of the shear modulus decreases more than when $k_{p,i}$ or $k_{a,i}$ are tunable DOF (see SI).

% \FloatBarrier

%new section for dos results
Since allowing shape indices as degrees of freedom shifts $p_0^*$, we expect that it also alters the vibrational density of states that describes the curvatures of the potential energy landscape in the rigid phase near $p_0^*$. As shown in the SI, double optimization on the $\{p_{0,i}\}$ DOFs reduces the curvatures and shifts the normal modes to lower frequencies. While previous work has suggested that additional signatures of double optimization, such as high-curvature directions in the cost function~\cite{bhaumik_mechanical_2023,stern_physical_2024}, can be found in eigenmodes of the cost Hessian, which in this case are identical to the vibrational normal modes since the cost function is simply the energy, we do not find any such signatures here. We conjecture that this is because the cost landscape and physical landscapes are already identical from the beginning of the double optimization process. As a result, there is no way in which double optimization can leave imprints on the energy landscape through coupling of two distinct landscapes.

Allowing $p_{0,i}$ as tunable DOF not only shifts $p_0^*$ but also increases the amount of structural order in the tissue (see SI). This ordering feature can be seen by sharper peaks in the pair correlation function. Consistent with this observation, we find a higher fraction of hexagonal cells $f_6$ when $\{p_{0,i}\}$ are added as new DOF (see SI). Importantly, the range of $f_6$, from 0.3 to 0.65, is tunable through adjustments in the mean and standard deviation of the $\{p_{0,i}\}$ distribution. This property can be used to mimic the level of hexagonal cells in epithelial tissues, which has been shown to change substantially between different stages of development \cite{zallen_cell-pattern_2004}.

\subsection{Non-monotonic relationship between rigidity shift and distribution width}

\begin{figure}
	\includegraphics[width=8cm,height=8cm,keepaspectratio]{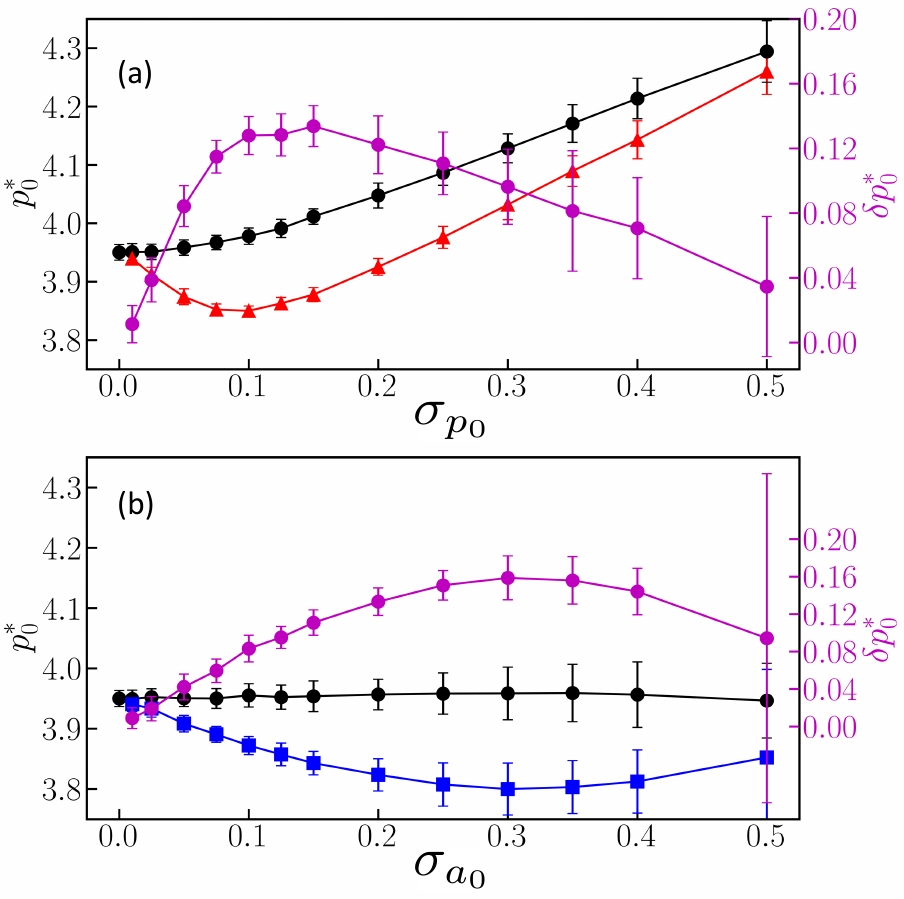}
	\caption{\label{fig:varying_sigma} The effect of polydispersity of tunable DOF distributions on the rigidity transition point. (a) The left axis shows the transition point $p_0^*$ versus the standard deviation $\sigma_{p_0}$ of the $\{p_{0,i}\}$ distribution. When only vertex positions can vary during energy minimization (black circles), $p_0^*$ increases with $\sigma_{p_0}$. However, when $\{p_{0,i}\}$ are also allowed to vary (red triangles), the behavior of $p_0^*$ versus $\sigma_{p_0}$ becomes non-monotonic. The right axis $\delta p^*_0$ shows the reduction of $p_0^*$ due to adding $\{p_{0,i}\}$ as degrees of freedom. (b) Same as (a), but with $\{a_{0,i}\}$ allowed to vary instead of $\{p_{0,i}\}$.}
\end{figure}

So far, we have used a fixed standard deviation ($\sigma = 0.2$) for the distribution of $\{p_{0,i}\}$. However, $\sigma$ significantly influences tissue rigidity \cite{li_mechanical_2019}, shifting $p_0^*$ upwards~\cite{li_mechanical_2019} (black circles in Fig. \ref{fig:varying_sigma}a). This raises the question: how does the shift in the transition $p_0^*$ due to adding $p_{0,i}$ as tunable DOF vary with $\sigma$? We observe a reduction in $p^*_0$ at all $\sigma$ (compare the red triangles to black circles in Fig. \ref{fig:varying_sigma}a). 
Interestingly, the magnitude of this reduction is non-monotonic. The purple curve in Fig. \ref{fig:varying_sigma}a, $\delta p^*_0$, shows that the shift in the transition is maximal at $\sigma \approx 0.15$. This suggests there is an optimal level of cell-to-cell fluctuations in biological tissues that enables double optimization to modulate rigidity.

To understand this non-monotonicity, we first note that as $\sigma$ approaches zero, the $\{p_{0,i}\}$ distribution approaches a delta-function and there are no tunable DOF. Therefore, $\delta p^*_0$ must increase away from that point. To understand why $\delta p^*_0$ decreases for $\sigma \gtrsim 0.15$, we analyzed the correlations between $p$ and $p_0$ across all cells, both with and without $p_0$ as tunable DOF. As expected, the Pearson's correlation coefficient $\rho(p,p_0)$ rises when we incorporate cell $p_0$ values as tunable DOF across all $\sigma$ values (see SI); the energy is lowered by bringing $p_i$ closer to $p_{0,i}$. But for $\sigma \gtrsim 0.15$, $p$ and $p_0$ already exhibit strong correlations even when $p_{0,i}$ are not tunable DOF. Introducing $p_{0,i}$ as tunable DOF only marginally enhances this correlation, so $\delta p^*_0$ decreases as shown in Fig.\ \ref{fig:varying_sigma}.
% \FloatBarrier
%

Finally, we consider the preferred cell areas, $A_{0,i}$. In tissues where $A_{0,i}=A_0$ is the same for all cells, altering $A_0$ while keeping $P_0$ fixed does not affect $p_0^*$ due to the confluency constraint ($\sum A_i = \mathrm{constant} = L^2$)~\cite{yang_correlating_2017, teomy_confluent_2018}. Yang et al.~\cite{yang_correlating_2017} found that the difference between $A_0$ and the actual area $\langle A \rangle = N/L^2$ alters the overall pressure of the system, but not the shear stresses. We find that even in the presence of heterogeneous $A_{0,i}$ values, $p_0^*$ is unaffected by changes in the \emph{average} target area $\langle A_0 \rangle$ (see SI), so in what follows we hold $\langle A_0 \rangle = 1$ fixed. The solid black circles in Fig.~\ \ref{fig:varying_sigma}b shows that varying the \emph{width} of the distribution of the dimensionless $a_{0,i}$ in Eq.~\ref{eq:energy_dimless} also does not affect the transition point. Note that defining target shape factors as $P_{0,i}/\langle A_i \rangle$ would introduce variability in the transition point with $\sigma_{a_0}$ (see SI). This suggests that the enhanced rigidity discussed in Ref. \cite{li_mechanical_2019} is caused by heterogeneity in target shape indices ($p_{0,i} = P_{0,i}/\sqrt{A_{0,i}}$) and not by heterogeneity in $P_{0,i}$.

Given these results, we promote $\{A_{0,i}\}$ to tunable DOF while keeping $\langle A_{0,i} \rangle$ fixed. This introduces \emph{two} sets of tunable DOF in the dimensionless energy in Eq.~\ref{eq:energy_dimless}, namely $\{a_{0,i}\}$ and $\{p_{0,i}\}$. We have already shown how introducing $\{p_{0,i}\}$ as tunable DOF affects rigidity, so now we consider the effects of $a_{0,i}$ in isolation.  To do so, we maintain a constant target shape factor for cells, i.e., $p_{0,i} = p_0$, by coupling the target perimeters $\{P_{0,i}\}$ with the target areas: $\{P_{0,i} = p_0 \sqrt{A_{0,i}}\}$. This allows us to consider only $\{a_{0,i}\}$ while keeping the average $\langle A_{0,i} \rangle = 1.0$ at homogeneous $p_{0,i} = p_0$. We find that introducing $\{a_{0,i}\}$ as tunable DOF leaves the scaling exponent for the shear modulus unchanged (see SI). Similar to $\{p_{0,i}\}$, the $\{a_{0,i}\}$ tunable DOF shift the transition downwards. This occurs at all values of the width of $\{a_{0,i}\}$ distribution, $\sigma_{a_0}$, with the maximum shift occurring around $\sigma_{a_0} \approx 0.3$. Correlations in $a_{0,i}$ from cell to cell causes $\tau_{ij}$ in Eq.~\ref{eq:tension} to vanish for some edges, shifting the percolation of nonzero tensions to lower $p_0^*$. 

\section{Discussion}

We have explored the effects of adding tunable degrees of freedom in 2D vertex models on the rigidity transition point, $p_0^*$. The transition is unaffected when cell stiffnesses $K_A$ and $K_P$ are allowed to vary. In contrast, introducing preferred cell areas or perimeters as tunable degrees of freedom significantly alters the tissue's energy landscape, shifting $p_0^*$ downwards. Learned spatial correlations in $p_0$ or $a_0$ can soften a tissue, and there are optimal values for the heterogeneity in $p_0$ ($\sigma_{p_0} \approx 0.15$) and $a_0$ ($\sigma_{a_0} \approx 0.3$) that lead to the largest shift of the transition.

tunable DOF have previously been introduced into networks that become rigid when the number of physical degrees of freedom equals the number of physical constraints~\cite{hagh_transient_2022,goodrich_principle_2015,hexner_role_2018,hexner_linking_2018}. In contrast, vertex models are highly under-constrained, and become rigid through geometric incompatibility~\cite{damavandi_energetic_2022}. Our finding that rigidity in these models is also strongly affected by tunable DOF suggests that vertex models can be used to study epithelial mechanics in terms of double optimization processes.

It is well-established that systems with fixed topology can learn intricate tasks \cite{stern_learning_2023}. While Hagh et al. \cite{hagh_transient_2022} demonstrated that jammed particle packings subject to frequent rearrangements can learn to identify ultra stable states, it is difficult to tune arbitrary mechanical responses into \emph{typical} jammed states because they are marginally stable to rearrangements~\cite{pisanty_private}. Confluent epithelial tissues lie in an intermediate state between these two extremes -- topological rearrangements, primarily in the form of T1 transitions, can occur but are not nearly as prevalent as in jammed packings. Our finding that preferred shape indices and cell areas effectively tune rigidity in vertex models suggests that introducing them as tunable DOF could be a fruitful way of obtaining complex responses in systems that allow topological rearrangements. 

The framework of physical learning with $\{p_{0,i}\}$ or $\{a_{0,i}\}$ as tunable DOF could provide a new paradigm for understanding biological tissue mechanics. Individual cells can control cell- and molecular-scale properties, including the concentration of adhesion molecules and myosin motors, which in turn govern the preferred shape index locally~\cite{wang_anisotropy_2020, sahu_small-scale_2020} and alter effective cell-cell interactions. Our work indicates that tissues should be able to learn if they follow a global gradient closely enough. In other systems, it has been possible to identify local learning rules that project sufficiently onto the global gradient to allow double optimization~\cite{pashine_directed_2019,hexner_effect_2020,hexner_periodic_2020,stern_supervised_2021, dillavou_demonstration_2022}. It would be interesting to study whether local rules governing the dynamics of cell shapes and tensions that have already been proposed~\cite{staddon_mechanosensitive_2019, noll_active_2017,claussen_geometric_2023} project onto gradients of useful global cost functions, or conversely, to hypothesize  cost functions for tissues and search for possible local learning rules that enable them to be minimized. More broadly, this framework could be useful for predicting how the dynamics of tissues arises from variation of cellular properties across developmental or evolutionary timescales.

\section*{Acknowledgments}
We thank Varda F. Hagh, R. Cameron Dennis, and Elizabeth Lawson-Keister for helpful discussions.  MLM and SA were supported by Simons Foundation \#454947, NSF-DMR-1951921 and SU's Orange Grid research computing cluster. IT and AJL were supported by NIH through Award 1-U01-CA-254886-01 and AJL by the Simons Foundation \#327939. AJL thanks CCB at the Flatiron Institute, a division of the Simons Foundation, and the Isaac Newton Institute for Mathematical Sciences at Cambridge University (EPSRC grant EP/R014601/1), for support and hospitality. IT acknowledges the CSIR-4PI institute for providing the HPC facility.

\section*{Data Availability}
The data that support the findings of this study
are available in an accompanying Dryad repository at [insert link].

%\bibliographystyle{unsrt}
%\bibliography{vertex_learning_refs}{}

\begin{thebibliography}{10}
	
	\bibitem{goodrich_principle_2015}
	Carl~P. Goodrich, Andrea~J. Liu, and Sidney~R. Nagel.
	\newblock The {Principle} of {Independent} {Bond}-{Level} {Response}: {Tuning}
	by {Pruning} to {Exploit} {Disorder} for {Global} {Behavior}.
	\newblock {\em Physical Review Letters}, 114(22):225501, June 2015.
	\newblock Publisher: American Physical Society.
	
	\bibitem{rocks_designing_2017}
	Jason~W. Rocks, Nidhi Pashine, Irmgard Bischofberger, Carl~P. Goodrich,
	Andrea~J. Liu, and Sidney~R. Nagel.
	\newblock Designing allostery-inspired response in mechanical networks.
	\newblock {\em Proceedings of the National Academy of Sciences},
	114(10):2520--2525, March 2017.
	\newblock Publisher: Proceedings of the National Academy of Sciences.
	
	\bibitem{hexner_role_2018}
	Daniel Hexner, Andrea~J. Liu, and Sidney~R. Nagel.
	\newblock Role of local response in manipulating the elastic properties of
	disordered solids by bond removal.
	\newblock {\em Soft Matter}, 14(2):312--318, 2018.
	
	\bibitem{hexner_linking_2018}
	Daniel Hexner, Andrea~J. Liu, and Sidney~R. Nagel.
	\newblock Linking microscopic and macroscopic response in disordered solids.
	\newblock {\em Physical Review E}, 97(6):063001, June 2018.
	
	\bibitem{rocks_limits_2019}
	Jason~W. Rocks, Henrik Ronellenfitsch, Andrea~J. Liu, Sidney~R. Nagel, and
	Eleni Katifori.
	\newblock Limits of multifunctionality in tunable networks.
	\newblock {\em Proceedings of the National Academy of Sciences},
	116(7):2506--2511, February 2019.
	\newblock Publisher: Proceedings of the National Academy of Sciences.
	
	\bibitem{pashine_directed_2019}
	Nidhi Pashine, Daniel Hexner, Andrea~J. Liu, and Sidney~R. Nagel.
	\newblock Directed aging, memory, and nature{\textquoteright}s greed.
	\newblock {\em Science Advances}, 5(12):eaax4215, December 2019.
	\newblock Publisher: American Association for the Advancement of Science.
	
	\bibitem{stern_learning_2023}
	Menachem Stern and Arvind Murugan.
	\newblock Learning {Without} {Neurons} in {Physical} {Systems}.
	\newblock {\em Annual Review of Condensed Matter Physics}, 14(1):417--441,
	2023.
	\newblock \_eprint: https://doi.org/10.1146/annurev-conmatphys-040821-113439.
	
	\bibitem{hexner_effect_2020}
	Daniel Hexner, Nidhi Pashine, Andrea~J. Liu, and Sidney~R. Nagel.
	\newblock Effect of directed aging on nonlinear elasticity and memory formation
	in a material.
	\newblock {\em Physical Review Research}, 2(4):043231, November 2020.
	\newblock Publisher: American Physical Society.
	
	\bibitem{hexner_periodic_2020}
	Daniel Hexner, Andrea~J. Liu, and Sidney~R. Nagel.
	\newblock Periodic training of creeping solids.
	\newblock {\em Proceedings of the National Academy of Sciences},
	117(50):31690--31695, December 2020.
	\newblock Publisher: Proceedings of the National Academy of Sciences.
	
	\bibitem{scellier_equilibrium_2017}
	Benjamin Scellier and Yoshua Bengio.
	\newblock Equilibrium {Propagation}: {Bridging} the {Gap} between
	{Energy}-{Based} {Models} and {Backpropagation}.
	\newblock {\em Frontiers in Computational Neuroscience}, 11, 2017.
	
	\bibitem{stern_supervised_2021}
	Menachem Stern, Daniel Hexner, Jason~W. Rocks, and Andrea~J. Liu.
	\newblock Supervised {Learning} in {Physical} {Networks}: {From} {Machine}
	{Learning} to {Learning} {Machines}.
	\newblock {\em Physical Review X}, 11(2):021045, May 2021.
	\newblock Publisher: American Physical Society.
	
	\bibitem{dillavou_demonstration_2022}
	Sam Dillavou, Menachem Stern, Andrea~J. Liu, and Douglas~J. Durian.
	\newblock Demonstration of {Decentralized} {Physics}-{Driven} {Learning}.
	\newblock {\em Physical Review Applied}, 18(1):014040, July 2022.
	\newblock Publisher: American Physical Society.
	
	\bibitem{wycoff_desynchronous_2022}
	J.~F. Wycoff, S.~Dillavou, M.~Stern, A.~J. Liu, and D.~J. Durian.
	\newblock Desynchronous learning in a physics-driven learning network.
	\newblock {\em The Journal of Chemical Physics}, 156(14):144903, April 2022.
	
	\bibitem{dillavou_machine_2023}
	Sam Dillavou, Benjamin~D. Beyer, Menachem Stern, Marc~Z. Miskin, Andrea~J. Liu,
	and Douglas~J. Durian.
	\newblock Machine {Learning} {Without} a {Processor}: {Emergent} {Learning} in
	a {Nonlinear} {Electronic} {Metamaterial}, November 2023.
	\newblock arXiv:2311.00537 [cond-mat].
	
	\bibitem{mongera_fluid--solid_2018}
	Alessandro Mongera, Payam Rowghanian, Hannah~J. Gustafson, Elijah Shelton,
	David~A. Kealhofer, Emmet~K. Carn, Friedhelm Serwane, Adam~A. Lucio, James
	Giammona, and Otger Camp{\`a}s.
	\newblock A fluid-to-solid jamming transition underlies vertebrate body axis
	elongation.
	\newblock {\em Nature}, 561(7723):401--405, September 2018.
	
	\bibitem{wang_anisotropy_2020}
	Xun Wang, Matthias Merkel, Leo~B. Sutter, Gonca Erdemci-Tandogan, M.~Lisa
	Manning, and Karen~E. Kasza.
	\newblock Anisotropy links cell shapes to tissue flow during convergent
	extension.
	\newblock {\em Proceedings of the National Academy of Sciences},
	117(24):13541--13551, June 2020.
	\newblock Publisher: National Academy of Sciences Section: Biological Sciences.
	
	\bibitem{claussen_geometric_2023}
	Nikolas~H. Claussen, Fridtjof Brauns, and Boris~I. Shraiman.
	\newblock A {Geometric} {Tension} {Dynamics} {Model} of {Epithelial}
	{Convergent} {Extension}, November 2023.
	\newblock arXiv:2311.16384 [cond-mat, physics:physics, q-bio].
	
	\bibitem{erdemci-tandogan_tissue_2018}
	Gonca Erdemci-Tandogan, Madeline~J. Clark, Jeffrey~D. Amack, and M.~Lisa
	Manning.
	\newblock Tissue {Flow} {Induces} {Cell} {Shape} {Changes} {During}
	{Organogenesis}.
	\newblock {\em Biophysical Journal}, 115(11):2259--2270, December 2018.
	
	\bibitem{sanematsu_3d_2021}
	Paula~C. Sanematsu, Gonca Erdemci-Tandogan, Himani Patel, Emma~M. Retzlaff,
	Jeffrey~D. Amack, and M.~Lisa Manning.
	\newblock {3D} viscoelastic drag forces contribute to cell shape changes during
	organogenesis in the zebrafish embryo.
	\newblock {\em Cells \& Development}, 168:203718, December 2021.
	
	\bibitem{hufnagel_mechanism_2007}
	Lars Hufnagel, Aurelio~A. Teleman, Herv{\'e} Rouault, Stephen~M. Cohen, and
	Boris~I. Shraiman.
	\newblock On the mechanism of wing size determination in fly development.
	\newblock {\em Proceedings of the National Academy of Sciences},
	104(10):3835--3840, March 2007.
	\newblock Publisher: Proceedings of the National Academy of Sciences.
	
	\bibitem{farhadifar_influence_2007}
	Reza Farhadifar, Jens-Christian R{\"o}per, Benoit Aigouy, Suzanne Eaton, and
	Frank J{\"u}licher.
	\newblock The {Influence} of {Cell} {Mechanics}, {Cell}-{Cell} {Interactions},
	and {Proliferation} on {Epithelial} {Packing}.
	\newblock {\em Current Biology}, 17(24):2095--2104, December 2007.
	
	\bibitem{park_unjamming_2015}
	Jin-Ah Park, Jae~Hun Kim, Dapeng Bi, Jennifer~A. Mitchel, Nader~Taheri Qazvini,
	Kelan Tantisira, Chan~Young Park, Maureen McGill, Sae-Hoon Kim, Bomi Gweon,
	Jacob Notbohm, Robert Steward~Jr, Stephanie Burger, Scott~H. Randell,
	Alvin~T. Kho, Dhananjay~T. Tambe, Corey Hardin, Stephanie~A. Shore, Elliot
	Israel, David~A. Weitz, Daniel~J. Tschumperlin, Elizabeth~P. Henske, Scott~T.
	Weiss, M.~Lisa Manning, James~P. Butler, Jeffrey~M. Drazen, and Jeffrey~J.
	Fredberg.
	\newblock Unjamming and cell shape in the asthmatic airway epithelium.
	\newblock {\em Nature Materials}, 14(10):1040--1048, October 2015.
	\newblock Number: 10 Publisher: Nature Publishing Group.
	
	\bibitem{bi_density-independent_2015}
	Dapeng Bi, J.~H. Lopez, J.~M. Schwarz, and M.~Lisa Manning.
	\newblock A density-independent rigidity transition in biological tissues.
	\newblock {\em Nature Physics}, 11(12):1074--1079, December 2015.
	\newblock Number: 12 Publisher: Nature Publishing Group.
	
	\bibitem{bi_motility-driven_2016}
	Dapeng Bi, Xingbo Yang, M.~Cristina Marchetti, and M.~Lisa Manning.
	\newblock Motility-{Driven} {Glass} and {Jamming} {Transitions} in {Biological}
	{Tissues}.
	\newblock {\em Physical Review X}, 6(2):021011, April 2016.
	\newblock Publisher: American Physical Society.
	
	\bibitem{chiang_glass_2016}
	M.~Chiang and D.~Marenduzzo.
	\newblock Glass transitions in the cellular {Potts} model.
	\newblock {\em Europhysics Letters}, 116(2):28009, December 2016.
	\newblock Publisher: EDP Sciences, IOP Publishing and Societ{\`a} Italiana di
	Fisica.
	
	\bibitem{devany_cell_2021}
	John Devany, Daniel~M. Sussman, Takaki Yamamoto, M.~Lisa Manning, and
	Margaret~L. Gardel.
	\newblock Cell cycle{\textendash}dependent active stress drives epithelia
	remodeling.
	\newblock {\em Proceedings of the National Academy of Sciences},
	118(10):e1917853118, March 2021.
	
	\bibitem{sahu_small-scale_2020}
	Preeti Sahu, Daniel M.~Sussman, Matthias R{\"u}bsam, Aaron F.~Mertz, Valerie
	Horsley, Eric R.~Dufresne, Carien M.~Niessen, M.~Cristina~Marchetti,
	M.~Lisa~Manning, and J.~M.~Schwarz.
	\newblock Small-scale demixing in confluent biological tissues.
	\newblock {\em Soft Matter}, 16(13):3325--3337, 2020.
	\newblock Publisher: Royal Society of Chemistry.
	
	\bibitem{wang_e-cadherin_2024}
	Xun Wang, Christian~M. Cupo, Sassan Ostvar, Andrew~D. Countryman, and Karen~E.
	Kasza.
	\newblock E-cadherin tunes tissue mechanical behavior before and during
	morphogenetic tissue flows, May 2024.
	\newblock Pages: 2024.05.07.592778 Section: New Results.
	
	\bibitem{tah_minimal_2025}
	Indrajit Tah, Daniel Haertter, Janice~M. Crawford, Daniel~P. Kiehart,
	Christoph~F. Schmidt, and Andrea~J. Liu.
	\newblock A minimal vertex model explains how the amnioserosa avoids
	fluidization during drosophila dorsal closure.
	\newblock 122(1):e2322732121.
	\newblock Publisher: Proceedings of the National Academy of Sciences.
	
	\bibitem{hagh_transient_2022}
	Varda~F. Hagh, Sidney~R. Nagel, Andrea~J. Liu, M.~Lisa Manning, and Eric~I.
	Corwin.
	\newblock Transient learning degrees of freedom for introducing function in
	materials.
	\newblock {\em Proceedings of the National Academy of Sciences},
	119(19):e2117622119, May 2022.
	\newblock Publisher: Proceedings of the National Academy of Sciences.
	
	\bibitem{damavandi_energetic_2022}
	Ojan~Khatib Damavandi, Varda~F. Hagh, Christian~D. Santangelo, and M.~Lisa
	Manning.
	\newblock Energetic rigidity. {I}. {A} unifying theory of mechanical stability.
	\newblock {\em Physical Review E}, 105(2):025003, February 2022.
	
	\bibitem{sharma_strain-controlled_2016}
	A.~Sharma, A.~J. Licup, K.~A. Jansen, R.~Rens, M.~Sheinman, G.~H. Koenderink,
	and F.~C. MacKintosh.
	\newblock Strain-controlled criticality governs the nonlinear mechanics of
	fibre networks.
	\newblock {\em Nature Physics}, 12(6):584--587, January 2016.
	
	\bibitem{shivers_scaling_2019}
	Jordan~L. Shivers, Sadjad Arzash, Abhinav Sharma, and Fred~C. MacKintosh.
	\newblock Scaling {Theory} for {Mechanical} {Critical} {Behavior} in {Fiber}
	{Networks}.
	\newblock {\em Physical Review Letters}, 122(18):188003, May 2019.
	
	\bibitem{merkel_minimal-length_2019}
	Matthias Merkel, Karsten Baumgarten, Brian~P. Tighe, and M.~Lisa Manning.
	\newblock A minimal-length approach unifies rigidity in underconstrained
	materials.
	\newblock {\em Proceedings of the National Academy of Sciences},
	116(14):6560--6568, April 2019.
	
	\bibitem{arzash_finite_2020}
	Sadjad Arzash, Jordan~L. Shivers, and Fred~C. MacKintosh.
	\newblock Finite size effects in critical fiber networks.
	\newblock {\em Soft Matter}, 16(29):6784--6793, July 2020.
	
	\bibitem{li_mechanical_2019}
	Xinzhi Li, Amit Das, and Dapeng Bi.
	\newblock Mechanical {Heterogeneity} in {Tissues} {Promotes} {Rigidity} and
	{Controls} {Cellular} {Invasion}.
	\newblock {\em Physical Review Letters}, 123(5):058101, July 2019.
	
	\bibitem{honda_description_1978}
	Hisao Honda.
	\newblock Description of cellular patterns by {Dirichlet} domains: {The}
	two-dimensional case.
	\newblock {\em Journal of Theoretical Biology}, 72(3):523--543, June 1978.
	
	\bibitem{sussman_cellgpu_2017}
	Daniel~M. Sussman.
	\newblock {cellGPU}: {Massively} parallel simulations of dynamic vertex models.
	\newblock {\em Computer Physics Communications}, 219:400--406, October 2017.
	
	\bibitem{bitzek_structural_2006}
	Erik Bitzek, Pekka Koskinen, Franz G{\"a}hler, Michael Moseler, and Peter
	Gumbsch.
	\newblock Structural {Relaxation} {Made} {Simple}.
	\newblock {\em Physical Review Letters}, 97(17):170201, October 2006.
	
	\bibitem{merkel_geometrically_2018}
	Matthias Merkel and M~Lisa Manning.
	\newblock A geometrically controlled rigidity transition in a model for
	confluent {3D} tissues.
	\newblock {\em New Journal of Physics}, 20(2):022002, February 2018.
	
	\bibitem{bhaumik_mechanical_2023}
	Himangsu Bhaumik and Daniel Hexner.
	\newblock Mechanical regularization, August 2023.
	\newblock arXiv:2308.05050 [cond-mat].
	
	\bibitem{stern_physical_2024}
	Menachem Stern, Andrea~J. Liu, and Vijay Balasubramanian.
	\newblock Physical effects of learning.
	\newblock {\em Physical Review E}, 109(2):024311, February 2024.
	\newblock Publisher: American Physical Society.
	
	\bibitem{zallen_cell-pattern_2004}
	Jennifer~A Zallen and Richard Zallen.
	\newblock Cell-pattern disordering during convergent extension in
	\textit{{Drosophila}}.
	\newblock {\em Journal of Physics: Condensed Matter}, 16(44):S5073--S5080,
	November 2004.
	
	\bibitem{yang_correlating_2017}
	Xingbo Yang, Dapeng Bi, Michael Czajkowski, Matthias Merkel, M.~Lisa Manning,
	and M.~Cristina Marchetti.
	\newblock Correlating cell shape and cellular stress in motile confluent
	tissues.
	\newblock {\em Proceedings of the National Academy of Sciences},
	114(48):12663--12668, November 2017.
	\newblock Publisher: National Academy of Sciences Section: Physical Sciences.
	
	\bibitem{teomy_confluent_2018}
	Eial Teomy, David~A. Kessler, and Herbert Levine.
	\newblock Confluent and nonconfluent phases in a model of cell tissue.
	\newblock {\em Physical Review E}, 98(4):042418, October 2018.
	\newblock Publisher: American Physical Society.
	
	\bibitem{pisanty_private}
	B.~Pisanty.
	\newblock (private communication, 2023).
	
	\bibitem{staddon_mechanosensitive_2019}
	Michael~F. Staddon, Kate~E. Cavanaugh, Edwin~M. Munro, Margaret~L. Gardel, and
	Shiladitya Banerjee.
	\newblock Mechanosensitive {Junction} {Remodeling} {Promotes} {Robust}
	{Epithelial} {Morphogenesis}.
	\newblock {\em Biophysical Journal}, 117(9):1739--1750, November 2019.
	
	\bibitem{noll_active_2017}
	Nicholas Noll, Madhav Mani, Idse Heemskerk, Sebastian~J. Streichan, and
	Boris~I. Shraiman.
	\newblock Active tension network model suggests an exotic mechanical state
	realized in epithelial tissues.
	\newblock {\em Nature Physics}, 13(12):1221--1226, December 2017.
	
\end{thebibliography}

\end{document}